\documentclass[twocolumn,showpacs,preprintnumbers]{revtex4}
\usepackage{amsmath}
\usepackage{graphicx}
\usepackage{dcolumn}
\usepackage{bm}
\usepackage{subfigure}
\usepackage{times}

\begin{document}

\title{Atom-photon, atom-atom and photon-photon entanglement
preparation\\
by fractional adiabatic passage}
\author{M. Amniat-Talab$^{1,2}$}
\email{amniyatm@u-bourgogne.fr}
\author{S. Gu\'{e}rin$^{1}$}
\email{sguerin@u-bourgogne.fr}
\author{N. Sangouard$^{1}$}
\author{H.R. Jauslin$^{1}$}
 \affiliation{$^{1}$Laboratoire de Physique, UMR CNRS 5027,
Universit\'{e} de Bourgogne, B.P. 47870, F-21078 Dijon,
France.\\$^{2}$Physics Department, Faculty of Sciences, Urmia
University, P.B. 165, Urmia, Iran.}

\date{\today }
\begin{abstract}
We propose a relatively robust scheme to generate maximally
entangled  states of (i) an atom and a cavity photon, (ii) two
atoms in their ground states, and (iii) two photons in two
spatially separate high-Q cavities. It is based on the interaction
via fractional adiabatic passage of a three-level  atom traveling
through a cavity mode and a laser beam. The presence of optical
phases  is emphasized.
\end{abstract}
 \maketitle
\section{Introduction\label{int}}
 One of the non-classical aspects of a quantum system made of $N$
parts is entanglement, for which the state vector of the system cannot be
written, in any basis, as a tensor product of independent substates. The
generation and the controlled manipulation of entangled states of $N$%
-particle systems is fundamental for the study of basic aspects of quantum
theory \cite{EPR,bell}. The idea is to apply a set of controlled coherent
interactions to the particles (atoms, ions, photons) of the system in order
to bring them into a tailored entangled state. The physics of entanglement
provides the basis of applications such as quantum information processing
and quantum communications. Very recently teleportation of quantum states
has been realized \cite{Riebe,Barrett} using atom-atom entanglement
following the proposal of Bennett \emph{et al.} \cite{teleport}. Particles
can then be viewed as carriers of quantum bits of information and the
realization of engineered entanglement is an essential ingredient of the
implementation of quantum gates \cite{qgate}.

Most experimental realizations of entanglement have been implemented with
photons. Although the individual polarization states of photons are easily
controlled, and their quantum coherence can be preserved over many
kilometers of an optical fiber \cite{tittel}, photons cannot be stored for
long times, and manipulations of collective entangled states present
considerable difficulties even when photons are confined in the same cavity.
The creation of long lived entangled pairs with atoms, on the other hand, is
a relatively recent pursuit which may provide reliable quantum information
storage. The entangled state of a pair of two-level atoms using pulse area
technique in a microwave cavity has been realized by Hagley \emph{et al.} %
\cite{haroche} based on the proposal of Cirac and Zoller \cite{cirac}.
However the pulse area technique is not robust with respect to the velocity
of the atoms and the exact-resonance condition. Recently a different scheme
has been proposed \cite{gong} to entangle two atoms using a tripod STIRAP
technique in a four-level atom-cavity-laser system in which one of the
pulses corresponds to the field of a cavity mode. Manipulation of
entanglement of two atoms in this scheme, however, requires to control a
geometric phase via an integral of Hamiltonian parameters over a closed path
in parameter space which is difficult in experimental implementations. The
generation of atom-photon entanglement has also been proposed in \cite{sun}
in a tripod-like laser-atom-cavity system which sustains two cavity modes.

In $\Lambda$-type systems, fractional STIRAP (f-STIRAP) is a
variation of STIRAP \cite{stirap} which allows the creation of any
preselected coherent superposition of the two degenerate ground
states \cite{vitanov}. As in STIRAP, the Stokes pulse linking the
initially unpopulated states $|e\rangle$ and $|g_{2}\rangle$,
arrives before the pump pulse linking the initially populated
state $|g_{1}\rangle$ to the excited state $|e\rangle$, but unlike
STIRAP where the Stokes pulse vanishes first, here the two pulses
vanish simultaneously while maintaining a constant finite ratio of
amplitudes. The f-STIRAP has been shown to increase the coherence
between the lower states of $\Lambda$-systems in nonlinear optics
experiments \cite{sautenkov}. The advantage of STIRAP is the
robustness of its control with respect to the precise tuning of
pulse areas, pulse delay, pulse widths, pulse shapes, and
detunings. Since f-STIRAP requires a precise ratio of pulse
endings, it is not as robust as STIRAP if two different pulses are
used. However in specific circumstances where a laser of elliptic
polarization can be used, f-STIRAP can be made as robust as STIRAP
\cite{vitanov}. In f-STIRAP as in STIRAP, if the evolution is
adiabatic (for instance with a slow transit of atoms across cw
fields), the dynamics of the system follows an
adiabatic dark state which does not involve the excited atomic state $%
|e\rangle$. Therefore this technique is immune to the detrimental
consequences of atomic spontaneous emission. The STIRAP technique has
interesting applications in the generation of coherent superposition of Fock
states \cite{parkins,parkins2} and of maximally polarization-entangled
photon states \cite{lange} in an optical cavity.

In this paper we consider neutral three-level $\Lambda$-type atoms
with two-fold degenerate
ground states $|g_{1}\rangle,|g_{2}\rangle$ and an excited state $|e\rangle$%
. The qubits are stored in the two ground states of the atoms. Our
scheme to create the entangled states is based on the resonant
interaction of the atoms with an optical cavity mode and a laser
field as follows:

(i) Atom-photon entanglement: the first atom initially in the ground state $%
|g_{1}\rangle$ interacts with the cavity mode (initially in the vacuum
state) and the laser field in the frame of f-STIRAP with the cavity-laser
sequence (meaning that the atom meets first the cavity).

(ii) Atom-atom entanglement: when the first atom has left the interaction
region, the second atom initially in the ground state $|g_{2}\rangle$
interacts with the laser-cavity sequence in the frame of STIRAP. After the
creation of an entangled state of the atoms, the cavity mode is left in the
vacuum state which is not entangled with the two atoms. Therefore the
decoherence effect of the cavity damping does not affect the atom-atom
entanglement before and after the interaction. The cavity damping must be
negligible only during the time of entanglement preparation.

(iii) Photon-photon entanglement: after the interaction of the
atom with the first cavity and the laser field in the frame of
f-STIRAP, the same atom interacts with the same laser field and
the second cavity in the frame of STIRAP. At the end of the
interaction, the atomic state factorizes and is left in the ground
state $|g_{2}\rangle$.
\section{Construction of the effective Hamiltonian}
\label{model} Figure 1 represents the linkage pattern of the
atom-cavity-laser system. The laser pulse associated to the Rabi frequency $%
\Omega (t)$ couples the states $|g_{1}\rangle $ and $|e\rangle$,
and the
cavity mode (Stokes pulse) with Rabi frequency $G(t)$ couples the states $%
|e\rangle $ and $|g_{2}\rangle $. The Rabi frequencies $\Omega(t)$
  and $G(t)$ are chosen real and positive without loss
of generality. These two fields interact with the atom with a time
delay, each of the fields is in one-photon resonance with the
respective transition. The semiclassical Hamiltonian (i.e. with a
classical laser field) of this system in the rotating-wave
approximation can be written in the atomic basis $\left\{
|g_{1}\rangle ,|e\rangle ,|g_{2}\rangle \right\} $
(in units of $\hbar $) as%
\begin{equation}
H(t)=\omega _{C}a^{\dag }a+\left[
\begin{array}{ccc}
0 & \Omega (t)e^{i(\omega _{L}t+\varphi _{L})} & 0 \\
\Omega (t)e^{-i(\omega _{L}t+\varphi _{L})} & \omega _{e} & G(t)a \\
0 & G(t)a^{\dagger } & 0%
\end{array}%
\right] ,
\end{equation}%
where  $a(a^{\dag })$ is the annihilation(creation) operator for
\begin{figure}[tbp]
\label{linkage}
\centerline{\subfigure{\includegraphics[width=4cm]{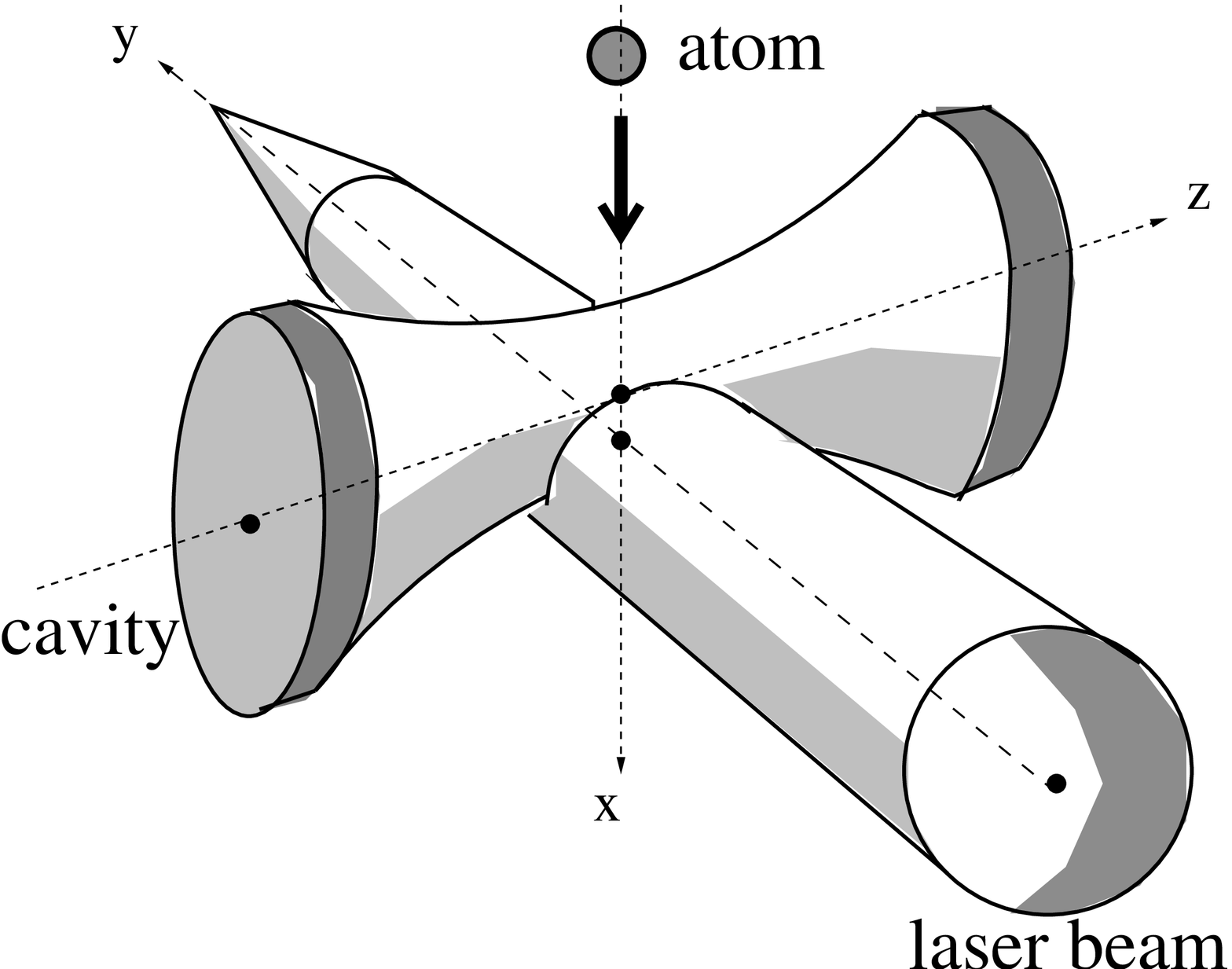}}
\subfigure{
\includegraphics[width=3cm]{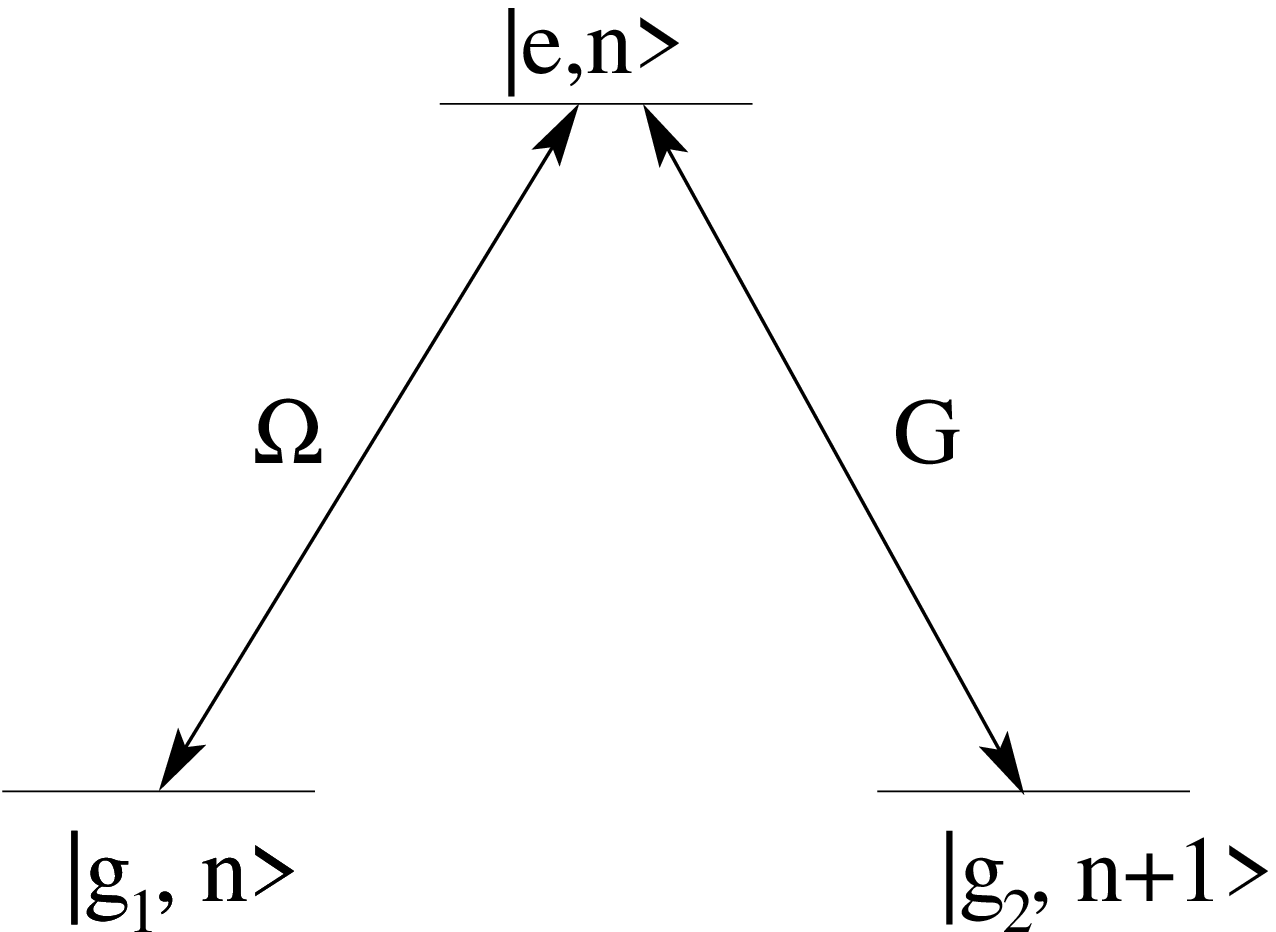}}}
\caption{ Experimental configuration and the linkage pattern of
atom-cavity-laser system with a two-photon resonance between
states $|g_{1},n\rangle $ and $|g_{2},n+1\rangle $.}
\end{figure}
the cavity mode, $\omega _{e}$ is the energy of the atomic excited
state $(\omega _{g_{1}}=\omega _{g_{2}}=0)$, $ \omega _{C},\omega
_{L}$ are the carrier frequencies of the cavity mode and the laser
field respectively $\omega _{C}=\omega _{L}=\omega _{e}$, and
$\varphi _{L}$ is the initial phase of the laser field. The
time-dependence of $\Omega (t)$ and $G(t)$ comes from the motion
of the atom across the laser and cavity fields and the time origin
is defined below.

The Hamiltonian $H(t)$ is block-diagonal in the manifolds $%
\{|g_{1},n\rangle ,|e,n\rangle ,|g_{2},n+1\rangle ;~n=0,1,2,...\}$, where $n$
is the number of photons in the cavity mode, $|e,n\rangle \equiv |e\rangle
\otimes |n\rangle $ and $|n\rangle $ is a $n$-photon Fock state. The vector $%
|g_{2},0\rangle $ is not coupled to any other ones, i.e.
$|g_{2},0\rangle $ is a \emph{stationary state} of the system. One
can thus restrict
the problem to the projection of the Hamiltonian in the subspace $%
\{|g_{1},0\rangle ,|e,0\rangle ,|g_{2},1\rangle \}:$
\begin{subequations}
\begin{eqnarray}
H_{P} &:=&PHP, \\
P &=&|g_{1},0\rangle \left\langle g_{1},0\right| +|e,0\rangle \left\langle
e,0\right| +|g_{2},1\rangle \left\langle g_{2},1\right| ,
\end{eqnarray}
if one considers the initial state $|g_{1},0\rangle $. The
associated dynamics is determined by the Schr\"{o}dinger equation $i\frac{%
\partial }{\partial t}|\Psi (t)\rangle =H_{P}(t)|\Psi (t)\rangle$.
The effective Hamiltonian can be written as
\end{subequations}
\begin{subequations}
\begin{eqnarray}
H^{\text{eff}} &=&R^{\dagger}H_{P}R-iR^{\dagger}\frac{\partial R}{\partial t}, \\
R(t) &=&|g_{1},0\rangle \langle g_{1},0|+e^{-i\omega_{L}
t}(|e,0\rangle \langle e,0|+|g_{2},1\rangle \langle
g_{2},1|),\nonumber\\
 \label{Rint}
\end{eqnarray}%
which reads in the basis $\left\{ |g_{1},0\rangle ,|e,0\rangle
,|g_{2},1\rangle\right\}$%
\end{subequations}
\begin{equation}
H^{\text{eff}}(t)=\left[
\begin{array}{ccc}
0 & \Omega (t)e^{i\varphi _{L}} & 0 \\
\Omega (t)e^{-i\varphi _{L}} & 0 & G(t) \\
0 & G(t) & 0%
\end{array}%
\right],
\end{equation}%
with the corresponding dynamics $i\frac{\partial }{\partial
t}|\Phi (t)\rangle =H^{\text{eff}}(t)|\Phi (t)\rangle $. The
relation between $|\Psi\rangle$ and $|\Phi\rangle$ is established
by unitary transformation $R$ as $|\Psi \rangle =R|\Phi\rangle$.
\section{Atom-photon entanglement}
\label{a-ph} The system is taken to be initially in the state $%
|g_{1},0\rangle $,
\begin{equation}
|\Phi (-\infty )\rangle =|g_{1},0\rangle =|\Psi (-\infty )\rangle
\label{inipsi}
\end{equation}
and we will transform it at the end of interaction into the
atom-photon entangled state
\begin{subequations}
\label{finphipsi}
\begin{eqnarray}
|\Phi (t &\rightarrow &+\infty )\rangle =\cos \vartheta |g_{1},0\rangle
-e^{-i\varphi _{L}}\sin \vartheta |g_{2},1\rangle , \\
|\Psi (t &\rightarrow &+\infty )\rangle =\cos \vartheta
|g_{1},0\rangle -e^{-i(\omega_{L} t+\varphi _{L})}\sin \vartheta
|g_{2},1\rangle,\nonumber\\
 \label{finpsi}
\end{eqnarray}
where $\vartheta $ is a constant mixing angle $(0\leq \vartheta \leq \pi /2)$%
. It is important to notice the presence of the generally unknown
absolute phase $\omega_{L} t+\varphi _{L}$  in the resulting
entangled state (\ref{finpsi}). This optical phase factor was not
taken into account by Parkins \emph{et al.} in the generation of
an arbitrary superpositions of Fock states \cite{parkins}. This
phase that changes rapidly as a function of the time, is expected
to be uncontrollable in practice.

 One of the instantaneous
eigenstates (the dark state) of $H^{\text{eff}}(t) $ which
corresponds to a zero eigenvalue and therefore to a zero dynamical
phase, is
\end{subequations}
\begin{equation}
|D(t)\rangle =\frac{1}{\sqrt{\Omega ^{2}(t)+G^{2}(t)}}[G(t)|g_{1},0\rangle
-\Omega (t)e^{-i\varphi _{L}}|g_{2},1\rangle ].  \label{dark}
\end{equation}%
As in f-STIRAP, the cavity-mode pulse comes first and is followed
after a certain time delay by the laser pulse, but the two pulses
vanish simultaneously, that can be asymptotically formulated as
\begin{equation}
\lim_{t\rightarrow -\infty }\frac{\Omega (t)}{G(t)}=0,\quad
\lim_{t\rightarrow +\infty }\frac{\Omega (t)}{G(t)}=\tan \vartheta .
\label{ratio}
\end{equation}%
The dark state (\ref{dark}) has consequently the limits
$|D(-\infty )\rangle =|g_{1},0\rangle $ and $|D(+\infty )\rangle
=\cos \vartheta |g_{1},0\rangle -e^{-i\varphi _{L}}\sin \vartheta
|g_{2},1\rangle $ with such a pulse sequence and allows thus one
to generate a coherent superposition of states by adiabatic
passage. It should be emphasized that this formulation in terms of
asymptotics (\ref{ratio}) does not describe correctly what occurs
at the beginning and the ending of f-STIRAP for a concrete
realization  which is not strictly adiabatic. In particular, using
this asymptotics would give a failure of f-STIRAP for Gaussian
pulses (considered below), which do not asymptotically give a
constant ending ratio for any delay and pulse width. The
inspection of the nonadiabatic couplings \cite{vitanov} shows that
what matters is that the Rabi frequency amplitudes end in a
constant ratio \emph{in a time interval where they are non
negligible}, and Eq. (\ref{ratio}) has to be understood in this
sense. The goal in the following is to show that such a pulse
sequence can be designed in a cavity by an appropriate choice of
the parameters.
\begin{figure}[tbp]
\includegraphics[width=8cm]{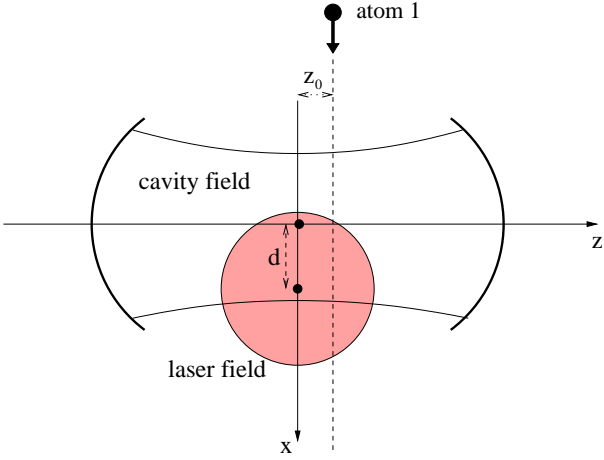}
\caption{ The geometry of the cavity-mode and the laser fields in the $xz$
plane with different waists ($W_{C}>W_{L}$), and the trajectory of the first
atom. The specific values of $z_{0}$ and $d$ are chosen such that the atom
interacts with the fields via f-STIRAP with the sequence cavity-laser. }
\label{geometry}
\end{figure}

In an optical cavity, the spatial variation of the atom-field
coupling for a Hermite-Gauss TEM$_{mn}$ mode is given by
\begin{eqnarray}
G_{mn}(x,y,z) &=&G_{0}~H_{m}\left( \frac{\sqrt{2}~x}{W_{C}}\right)
~H_{n}\left( \frac{\sqrt{2}~y}{W_{C}}\right)   \notag  \label{Gmn} \\
&\times &e^{-(x^{2}+y^{2})/W_{C}^{2}}~~\cos \left( \frac{2\pi z}{\lambda }%
\right),
\end{eqnarray}%
where $G_{0}=\mu\sqrt{\omega_{C}/(2\epsilon_{0}V_{\text{mode}})}$
with $\mu$, $V_{\text{mode}}$  respectively the dipole moment of
the atomic transition and the effective
volume of the cavity mode. The transverse distribution is determined by Hermite polynomials $H_{m}$, $%
H_{n}$ and the cavity waist $W_{C}$ \cite{keller}. The standing wave along
the cavity z-axis gives rise to a $\cos (2\pi z/\lambda )$ dependence of the
mode with the wavelength $\lambda $. A particular transverse mode is
selected by adjusting the cavity length. We consider the maximum coupling
mode TEM$_{00}$ resonant with the $|e\rangle \leftrightarrow |g_{2}\rangle $
transition of the atom
\begin{equation}
G(x,y,z)=G_{0}e^{-(x^{2}+y^{2})/W_{C}^{2}}\cos \left( \frac{2\pi z}{\lambda }%
\right) .  \label{G00}
\end{equation}%
\newline
Figure 2 shows a situation where an atom initially in the state $%
|g_{1}\rangle $ falls with velocity $v$ (on the $y=0$ plane and
$z=z_{0}$ line) through an optical cavity initially in the vacuum
state $|0\rangle $ and then encounters the laser beam, which is
parallel to the y axis (orthogonal to the cavity axis and the
trajectory of the atom). The laser beam of waist $W_{L}$ is
resonant with the $|e\rangle \leftrightarrow |g_{1}\rangle $
transition. The distance between center of the cavity and the
laser axis is $d$. The traveling atom encounters the time
dependent and delayed Rabi frequencies of the cavity and the laser
fields as follows
\begin{subequations}
\begin{eqnarray}
G(t) &=&G_{0}~e^{-(vt)^{2}/W_{C}^{2}}\cos \left( \frac{2\pi z_{0}}{\lambda }%
\right) , \\
\Omega (t) &=&\Omega _{0}~e^{-z_{0}^{2}/W_{L}^{2}}~e^{-(vt-d)^{2}/W_{L}^{2}},
\end{eqnarray}%
where the time origin is defined when the atom meets the center of
the cavity $x=0$. The appropriate values of $z_{0}$ and $d$ that
lead to the f-STIRAP process can be extracted from a contour plot
of the final population $P_{|g_{1},0\rangle }:=|\langle
g_{1},0|\Phi (+\infty )\rangle |^{2}$ as a function of $z_{0}$ and
$d$ that
we calculate numerically (see Fig. \ref{contour}). The white dot in Fig. \ref%
{contour} shows values of $z_{0}$ and $d$ to obtain a f-STIRAP process with $%
\vartheta \simeq \pi /4$ (called half-STIRAP). It has been chosen
such that
at the end of interaction $P_{|g_{1},0\rangle }\simeq P_{|g_{2},1\rangle }\simeq 0.5$ and $%
P_{|e,0\rangle }\simeq 0.$
\begin{figure}[tbp]
\includegraphics[width=8cm]{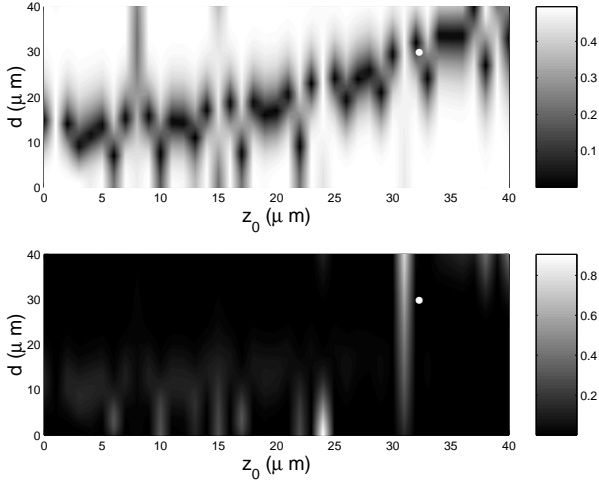}
\caption{Top panel: contour plot of the final population $|\frac{1}{2}%
-P_{|g_{1},0\rangle }|$ as a function of $z_{0}$ and $d$ (black areas
correspond to approximately half population transfer) with the pulse
parameters as $W_{L}=20\protect\mu $m, $~W_{C}=30\protect\mu $m,$~v=2$ m/s,$~%
\protect\lambda =780$nm, $~\Omega _{0}=50\frac{v}{W_{L}},~G_{0}=50\frac{v}{%
W_{C}}$. Bottom panel: the same plot for the population of the
excited state $P_{|e,0\rangle }:=|\langle e,0|\Phi (+\infty
)\rangle |^{2}$ where black areas correspond to approximately zero
population transfer. The white dot shows specific values of
$z_{0}$ and $d$ used in Fig. \ref{Fstirap} to obtain a half-STIRAP
process.} \label{contour}
\end{figure}
\begin{figure}[tbp]
\includegraphics[width=8cm]{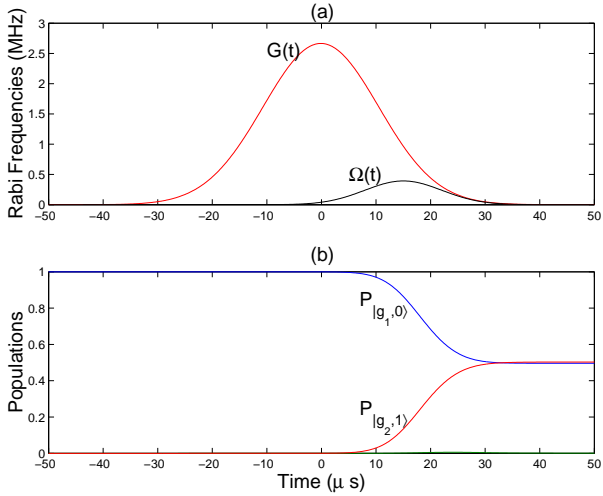}
\caption{(a) Rabi frequencies of the cavity-mode and the laser
field for the first atom corresponding to the same pulse
parameters and the specific values $z_{0}=31.9\protect\mu $m,
$d=30.2\protect\mu $m of the white dot in Fig. \ref{contour}. (b)
Time evolution of the populations for the trajectory of the first
atom which represents a half-STIRAP.} \label{Fstirap}
\end{figure}

Figure \ref{Fstirap} shows (a) the cavity-laser pulse sequence of
half-STIRAP for the first atom, and (b) the time evolution of
populations which shows half-half population for the states
$|g_{1},0\rangle ,|g_{2},1\rangle $ and zero population for the
state $|e,0\rangle $ at the end of the interaction. This case
corresponds to the generation of the maximally atom-photon
entangled state $1/\sqrt{2}\big(|g_{1},0\rangle-e^{-i(\omega_{L}
t+\varphi _{L})}|g_{2},1\rangle\big)$ by adiabatic passage.
Assuming Gaussian pulse profiles for $\Omega (t)$ and $G(t)$ of widths $%
T_{L}=W_{L}/v$ and $T_{C}=W_{C}/v$ respectively, we have the
sufficient condition of adiabaticity \cite{vitanov}:
\end{subequations}
\begin{equation}
\Omega _{0}T_{L},G_{0}T_{C}\gg 1.
\end{equation}%

We remark that the case $\vartheta =\pi /2$ corresponds to the
standard STIRAP with the final state $|\Psi (t\rightarrow +\infty
)\rangle =-e^{-i(\omega_{L} t+\varphi _{L})}|g_{2},1\rangle $,
i.e. to the generation of a single-photon Fock state in the cavity
mode without population transfer to the atomic excited state at
the end of interaction. Here the optical phase factor appears as
an irrelevant global phase factor. A one-photon Fock state has
been produced in such a way in an optical cavity via STIRAP by
Hennrich \emph{et al.} \cite{hennrich} based on the proposal of
Refs. \cite{parkins,kuhn}. A robust scheme for the generation of
multi-photon Fock states in a microwave cavity via bichromatic
adiabatic passage has been proposed in Ref. \cite{amniat}.
\section{atom-atom entanglement\label{a-a}}
\begin{figure}[h]
\includegraphics[width=6cm]{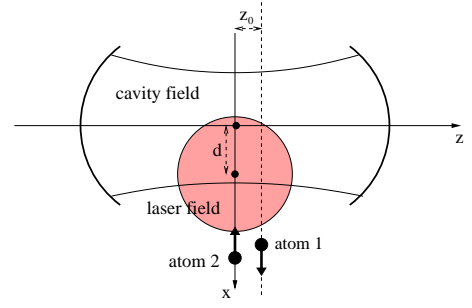}
\caption{ The proposed geometry of the cavity and the laser fields
in $xz$ plane as well as the trajectory of the atoms for
generation of atom-atom entanglement. The second atom initially in the ground state $%
|g_{2}\rangle$ arrives at the center of the cavity with a time
delay $\tau$. This atom encounters the sequence laser-cavity on
the line $z=0$.} \label{atom2}
\end{figure}
In this section we consider a situation where the first atom has been
entangled with the cavity mode via f-STIRAP as described by Eq. (%
\ref{finpsi}), and the second atom initially in the ground state $%
|g_{2}^{(2)}\rangle $ is going to interact with the \emph{same} laser and
cavity-mode fields but through a STIRAP process (see Fig. \ref%
{atom2}). The superscript labels the two atoms. The state of the atom%
$^{(2)}$-atom$^{(1)}$-cavity system after entanglement of the
atom$^{(1)}$ reads
\begin{equation}
|\Psi (t)\rangle =|g_{2}^{(2)}\rangle \big(\cos \vartheta
|g_{1}^{(1)},0\rangle -e^{-i(\omega_{L} t+\varphi _{L})}\sin
\vartheta |g_{2}^{(1)},1\rangle \big).
\end{equation}%
The second atom moves on the line $z=0$ in the same plane (see
Fig. \ref{atom2}) as the first one such that the two atoms
experience the same optical phase $e^{i\omega t}$ of the laser
field. It encounters time-dependent and delayed Rabi frequencies
given by
\begin{subequations}
\begin{eqnarray}
G^{(2)}(t) &=&G_{0}e^{-\left[ v(t-\tau )\right] ^{2}/W_{C}^{2}}, \\
\Omega ^{(2)}(t) &=&\Omega _{0}e^{-\left[ v(t-\tau )+d\right]
^{2}/W_{L}^{2}},
\end{eqnarray}%
where $\tau $ is the time delay between the two atoms. By standard
STIRAP, with the sequence of laser-cavity (see Fig. 6), we can
transfer the population from the initial state
$|g_{2}^{(2)},1\rangle$ to the final state
$|g_{1}^{(2)},0\rangle$. On the other hand the state
$|g_{2}^{(2)},0\rangle$ is stationary with respect to this STIRAP
process. Using the transformation
$R^{(2)}(t)=|g_{1}^{(2)},0\rangle \langle
g_{1}^{(2)},0|+e^{-i\omega_{L}t}\big(|e^{(2)},0\rangle \langle
e^{(2)},0|+|g_{2}^{(2)},1\rangle \langle g_{2}^{(2)},1|\big)$,
this results in
\end{subequations}
\begin{equation}
|g_{2}^{(2)},0\rangle \rightarrow |g_{2}^{(2)},0\rangle ,\quad
-e^{-i(\omega_{L}t+\varphi _{L})}|g_{2}^{(2)},1\rangle \rightarrow
|g_{1}^{(2)},0\rangle . \label{evo-dark}
\end{equation}%
Hence, if the second atom encounters the laser field before the cavity field
in the frame of a standard STIRAP, the final state of the atom$^{(2)}$-atom$%
^{(1)}$-cavity system will be
\begin{eqnarray}
|\Psi (+\infty )\rangle  &=&\cos \vartheta |g_{2}^{(2)},0\rangle
|g_{1}^{(1)}\rangle +\sin \vartheta
|g_{1}^{(2)},0\rangle |g_{2}^{(1)}\rangle   \notag \\
&=&|0\rangle \big(\cos \vartheta |g_{2}^{(2)}\rangle
|g_{1}^{(1)}\rangle +\sin \vartheta
|g_{1}^{(2)}\rangle |g_{2}^{(1)}\rangle \big).\nonumber\\
\label{entang12}
\end{eqnarray}%
Since the cavity-mode state factorizes and is left in the vacuum state,
there is no projection noise when one traces over the unobserved cavity
field, and the cavity is ready to prepare another entanglement. We can
manipulate this entanglement coherently to reach the maximal atom-atom
entanglement by tuning the ratio of fields such that $\tan \vartheta =1$ in
the f-STIRAP stage, as shown in Figs. 3 and 4.
\begin{figure}[tbp]
\includegraphics[width=7cm]{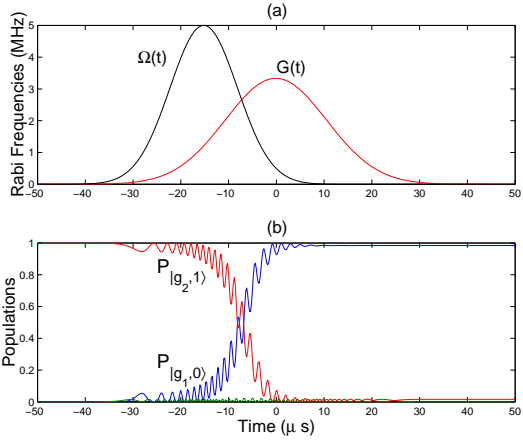}
\caption{ (a) Rabi frequencies of the fields for the second atom travelling
on the line $z=0$ with the same parameters of Fig. \ref{Fstirap}. (b) Time
evolution of the populations.}
\label{pop-atom2}
\end{figure}

Figure \ref{pop-atom2} shows the successful STIRAP process for the
second atom (moving along the line $z=0$) that allows one to
generate the entangled state (\ref{entang12}).

The generation of atom-atom entanglement with the two atoms
interacting simultaneously ($\tau=0$ in Fig. 5) with the cavity
mode, that can be described by a two-atom dark state presented in
Ref. \cite{pellizari}, will be discussed elsewhere.
\section{photon-photon entanglement\label{ph-ph}}
 In Refs. \cite{gerry,bergou,plenio} among many others, different schemes have been
proposed to entangle two and three microwave cavities through the
interaction with a Rydberg atom. A method of generating particular entangled
states of two cavities appeared as an intermediate step in the teleportation
procedure proposed by Davidovich \emph{et al.} \cite{davidovich}. Here we
propose another scheme to entangle two optical cavities interacting with an
atom. We consider a situation where an atom has been entangled with the
first single-mode cavity via a f-STIRAP technique as described in section %
\ref{model}, and it interacts next with another single-mode optical cavity,
initially in the vacuum state, and the same laser field (see Fig. \ref%
{2cavity}).  The distance $z_{0}$ between the axis of motion of
the atom and the center of the first cavity, ensures having an
f-STIRAP process for the
first cavity. Hence the  state of the cavity$^{(2)}$-atom-cavity$%
^{(1)}$ system after the f-STIRAP process is
\begin{equation}
|\Psi (t)\rangle =|0^{(2)}\rangle \big(\cos \vartheta
|g_{1},0^{(1)}\rangle -e^{-i(\omega _{L}t+\varphi _{L})}\sin
\vartheta |g_{2},1^{(1)}\rangle \big),
\end{equation}
\begin{figure}[tbp]
\includegraphics[width=8cm]{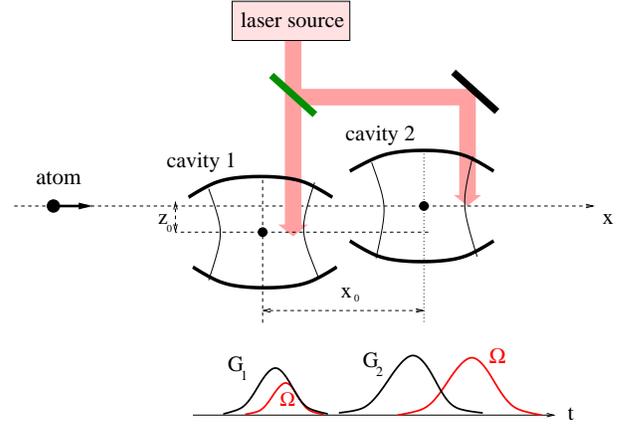}
\caption{Proposed setup for entangling of two photons generated in
two optical cavities interacting with a three-level atom and a
laser field, and the schematic pulse sequences. The atom interacts
with the first cavity via f-STIRAP in the same conditions of Fig.
\ref{geometry} and next with the second cavity via standard STIRAP
with the sequence cavity-laser in each one. The propagation
direction of the laser beams are perpendicular to the plane xz as
the figure 1.} \label{2cavity}
\end{figure}
where the superscripts denote the number of cavities. If the atom interacts
with the second cavity in the frame of a standard STIRAP and the atom
encounters the cavity mode before the laser pulse, since the state $%
|g_{1},0^{(2)}\rangle $ evolves to $|g_{2},1^{(2)}\rangle $ and the state $%
|g_{2},0^{(2)}\rangle $ does not change during the interaction:
\begin{eqnarray}
|g_{1},0^{(2)}\rangle &\rightarrow&-e^{-i(\omega
_{L}t+\varphi _{L})}|g_{2},1^{(2)}\rangle ,\nonumber\\
|g_{2},0^{(2)}\rangle &\rightarrow&|g_{2},0^{(2)}\rangle ,
\label{evo-dark3}
\end{eqnarray}%
the final state of the system will be (up to an
irrelevant common phase factor)
\begin{equation}
|\Psi (+\infty )\rangle =|g_{2}\rangle \big(\cos \vartheta
|1^{(2)},0^{(1)}\rangle +e^{i\alpha}\sin \vartheta
|0^{(2)},1^{(1)}\rangle \big), \label{entangcav}
\end{equation}%
where $\alpha=2\pi(x_{0}^{2}+z_{0}^{2})^{1/2}/\lambda$ is the
phase shift of the laser field due to the optical path difference
between the two cavities. Since the atomic state factorizes and is
left in the ground state $|g_{2}\rangle $, the atom does not have
spontaneous emission and it could be used to prepare another
entanglement.
\section{Discussions and Conclusions}
We have proposed a robust scheme to generate atom-photon,
atom-atom and photon-photon entanglement, using a combination of
f-STIRAP and STIRAP techniques in $\Lambda$-systems. This scheme
is robust with respect to  variations of the velocity of the atom
$v$, of the peak Rabi frequencies $G_{0},\Omega_{0}$ and  of the
pulse detunings, but not with respect to the parameters $d,z_{0}$.
For given values of $W_{C},W_{L}$, the adapted values of $d$ and
$z_{0}$ in the f-STIRAP process can be determined from a contour
plot of the final populations as explained in section \ref{a-ph}.

The presence of optical phases in the case of atom-photon
entanglement, expected to be uncontrollable, was emphasized.  In
the case of atom-atom entanglement the optical phase is not
present, as long as the two atoms move in the same plane
perpendicular to the propagation direction of the laser beam.

Dissipation in the form of spontaneous emission and cavity damping
is another important practical issue. The adiabatic passage
technique is robust against the effects of spontaneous emission,
as the excited atomic state is never appreciably populated. Cavity
damping is certainly a problem as its effects come into play as
soon as the cavity mode is excited, leading to a degradation of
the adiabatic transfer. In this analysis we have assumed that the
 interaction time between the  atom and the fields $T_{\text{int}}\approx W_{C}/v\approx W_{L}/v$ is
 short compared to the
 cavity lifetime $T_{\text{cav}}$,  which are
  essential for an experimental realization.

 Since the decay rate of the cavity  scales
 with the number of photons present in the cavity ($T_{\text{int}}\ll
 T_{\text{cav}}/n$), our
 scheme involving only one cavity photon requires $T_{\text{int}}\ll
 T_{\text{cav}}$. In a real experiment, it is desirable that
 the entangled states are as long-lived as possible. This requires
 in the optical domain, where $T_{\text{int}}\approx$ 15 $\mu$s, a
 cavity lifetime of $T_{\text{cav}}\gg 15 \mu$s. This is beyond the currently available optical cavities
 where $T_{\text{cav}}\approx$1 $\mu$s. One could still consider  the generation of atom-atom
 entanglement in an optical cavity using a two-atom dark state of the type presented in
 Ref. \cite{pellizari} which does  not require such a stringent  constraint
 for the cavity lifetime. In the microwave domain,
cavities with a photon lifetime of 1 ms
  \cite{raimondRMP} and of 0.3 s  \cite{waltherPRL} have been made. The  upper limit
 of interaction time is  $T_{\text{int}}=100$ $\mu$s (atom with a velocity
 of 100 m/s with the cavity mode waist of $W_{C}=6$ mm). The condition of global adiabaticity
$G_{0}~T_{\text{int}}\gg 1$ for the typical value of
$G_{0}\approx0.15$ MHz \cite{raimondRMP} is
 well satisfied $G_{0}~T_{\text{int}}\approx 15$. The proposed schemes of  entanglement generation
 could be implemented in a microwave cavity by
 using a maser field  and atomic
 Rydberg states.
\begin{acknowledgments}
M. A-T. wishes to acknowledge the financial support of the MSRT of
Iran and SFERE. We acknowledge support from the Conseil
R\'{e}gional de Bourgogne.
\end{acknowledgments}


\end{document}